\documentstyle[frascatiphys,epsfig]{article}

\begin{document}

\begin{flushright}
\underline{\bf Preprint LMU-00-05}
\end{flushright}

\vspace{0.2cm}

\begin{center}
{\Large\bf Neutrinos: How Do They Mix and Violate $CP$ ? \footnote{
Invited talk given by one of us (H.F.) at the 14th Rencontres de Physique
de la Vallee d'Aoste, La Thuile, Aosta Valley, Italy, February 2000}}
\end{center}

\vspace{0.5cm}
\begin{center}
{\bf Harald Fritzsch} \footnote{Electronic address:
bm@theorie.physik.uni-muenchen.de}
~ and ~ {\bf Zhi-zhong Xing} \\
{\sl Sektion Physik, Universit$\sl\ddot{a}$t M$\sl\ddot{u}$nchen,
Theresienstrasse 37A, 80333 Munich, Germany}
\end{center}

\vspace{3cm}
\begin{abstract}
We discuss a simple model of lepton mixing and $CP$
violation based on the flavor democracy of charge leptons and the
mass degeneracy of neutrinos. A
nearly bi-maximal flavor mixing pattern, which is favored by current
data on atmospheric and solar neutrino oscillations, emerges naturally
from this model after explicit symmetry breaking.
The rephasing-invariant strength of $CP$ or $T$ violation
can be as large as one percent, leading to significant
probability asymmetries between $\nu_\mu \rightarrow \nu_e$
and $\bar{\nu}_\mu \rightarrow \bar{\nu}_e$ (or
$\nu_e \rightarrow \nu_\mu$) transitions in the long-baseline
neutrino experiments. The possible matter effects on $CP$- and
$T$-violating asymmetries are also taken into account.
\end{abstract}

\newpage

Since its foundation in the 1960's the standard electroweak model,
which unifies the weak and electromagnetic interactions, has
passed all experimental tests. Neither significant evidence for the
departures from the standard model nor convincing hints for the
presence of new physics has been found thus far at HERA, LEP, SLC,
Tevatron
and other high-energy facilities. In spite of the impressive
success of the standard model, many physicists believe that it
does not represent the final theory, but serves merely as an effective
theory originating from a more fundamental, yet unknown theoretical
framework.
For instance there is little understanding, within the standard model,
about the intrinsic physics of the electroweak symmetry breaking,
the hierarchy of charged fermion mass spectra, the vanishing or
smallness
of neutrino masses, and the origin of flavor mixing and $CP$ violation.
Any attempt towards gaining an insight into such problems
inevitably requires significant steps to go beyond the standard model.

The investigations of fermion mass generation and flavor mixing
problems, which constitute an important part of today's particle
physics, can be traced back to the early 1970's, soon after the
establishment of the standard electroweak model. Since then many
different models or approaches have been 
developed\cite{FXReview}. From the theoretical point of
view, however, our understanding of the fermion mass spectrum
remains quite unsatisfactory. Before a significant breakthrough
takes place on the theoretical side, the phenomenological
approaches will remain to play a crucial role in interpreting new
experimental data on quark mixing, $CP$ violation, and neutrino
oscillations. They are expected to provide useful hints towards
discovering the full dynamics of fermion mass generation and
$CP$ violation.

In this talk I shall concentrate on neutrino oscillations, the
leptonic counterparts of the flavor mixing phenomena for quarks,
and study in particular the interesting prospects of
finding $CP$ violation in neutrino oscillations.

The recent observation of atmospheric and solar neutrino
anomalies, in particular by the Super-Kamiokande experiment\cite{SK},
has provided a strong indication that neutrinos are massive and
lepton flavors are mixed. As there exist at least three
different lepton families, the flavor mixing matrix may
in general contain non-trivial complex phase terms. Hence $CP$
or $T$ violation is naturally expected in the lepton sector.

A violation of $CP$ invariance in the quark sector can result in
a variety of observable effects in hadronic weak decays.
Similarly $CP$ or $T$
violation in the lepton sector can manifest itself
in neutrino oscillations\cite{Cabibbo}.
The best and probably the only way to observe $CP$- or
$T$-violating effects in neutrino oscillations is to carry out
the long-baseline appearance neutrino experiments\cite{Factory}.

In the scheme of three lepton families, the $3\times 3$ flavor
mixing matrix $V$ links the neutrino mass eigenstates
$(\nu_1, \nu_2, \nu_3)$ to the neutrino flavor eigenstates
$(\nu_e, \nu_\mu, \nu_\tau)$:
\begin{equation}
\left ( \matrix{
\nu_e \cr
\nu_\mu \cr
\nu_\tau \cr} \right ) \; =\;
\left ( \matrix{
V_{e1}  & V_{e2}        & V_{e3} \cr
V_{\mu 1}       & V_{\mu 2}     & V_{\mu 3} \cr
V_{\tau 1}      & V_{\tau 2}    & V_{\tau 3} \cr} \right )
\left ( \matrix{
\nu_1 \cr
\nu_2 \cr
\nu_3 \cr} \right ) \; .
\end{equation}
If neutrinos are massive Dirac fermions, $V$ can be parametrized
in terms of three rotation angles and one $CP$-violating phase.
If neutrinos are Majorana fermions, however, two additional
$CP$-violating phases are in general needed to fully parametrize $V$.
The strength of $CP$ violation in neutrino oscillations, no
matter whether neutrinos are of the Dirac or Majorana type,
depends only upon a universal parameter $\cal J$, which
is defined through
\begin{equation}
{\rm Im} \left (V_{il}V_{jm} V^*_{im}V^*_{jl} \right )
\; =\; {\cal J} \sum_{k,n} \epsilon^{~}_{ijk} \epsilon^{~}_{lmn}
\; .
\end{equation}
The asymmetry between the probabilities of two $CP$-conjugate
neutrino transitions, due to the $CPT$
invariance and the unitarity of $V$, is uniquely given as
\begin{eqnarray}
\Delta_{CP} & = & P(\nu_\alpha \rightarrow \nu_\beta) -
P(\bar{\nu}_\alpha
\rightarrow \bar{\nu}_\beta) \; \nonumber \\
& = & -16 {\cal J} \sin F_{12} \sin F_{23} \sin F_{31} \;
\end{eqnarray}
with $(\alpha, \beta) = (e,\mu)$, $(\mu, \tau)$ or $(\tau, e)$,
$F_{ij} = 1.27 \Delta m^2_{ij} L/E$ and
$\Delta m^2_{ij} = m^2_i -m^2_j$, in which $L$ is the distance
between the neutrino source and the detector
(in unit of km) and $E$ denotes the neutrino beam energy (in unit of
GeV). The $T$-violating asymmetry can be obtained in a
similar way:
\begin{eqnarray}
\Delta_T & = & P(\nu_\alpha \rightarrow \nu_\beta)
- P(\nu_\beta \rightarrow \nu_\alpha) \; 
\nonumber \\
& = & P(\bar{\nu}_\beta \rightarrow \bar{\nu}_\alpha)
- P(\bar{\nu}_\alpha \rightarrow \bar{\nu}_\beta) \; 
\nonumber \\
& = & -16 {\cal J} \sin F_{12} \sin F_{23} \sin F_{31} \; .
\end{eqnarray}
These formulas show clearly that $CP$ or $T$ violation is a feature
of all three lepton families. The relationship $\Delta_T
= \Delta_{CP}$ is a straightforward consequence
of $CPT$ invariance.
The observation of $\Delta_T$ might basically be free from the
matter effects of the earth, which is possible to fake the genuine
$CP$ asymmetry $\Delta_{CP}$ in any long-baseline neutrino
experiment. The joint measurement of $\nu_\alpha \rightarrow
\nu_\beta$ and $\nu_\beta \rightarrow \nu_\alpha$ transitions to
determine $\Delta_T$ is, however, a challenging task in practice.
Probably it could only be carried out in a neutrino factory,
whereby high-quality neutrino beams can be produced with high-intensity
muon storage rings\cite{Factory}.

The analyses of current data\cite{SK,CHOOZ}
yield $\Delta m^2_{\rm sun} \ll \Delta m^2_{\rm atm}$ as well as
$|V_{e3}|^2 \ll 1$, implying that the atmospheric and solar
neutrino oscillations are approximately decoupled.
A reasonable interpretation of those data follows from setting
$\Delta m^2_{\rm sun} = |\Delta m^2_{12}|$ and
$\Delta m^2_{\rm atm} = |\Delta m^2_{23}| \approx |\Delta m^2_{31}|$.
In this approximation $F_{31} \approx -F_{23}$ holds.
The $CP$- and $T$-violating asymmetries can then be
simplified as
\begin{equation}
\Delta_{CP} \; =\; \Delta_T \; \approx \;
16 {\cal J} \sin F_{12} \sin^2 F_{23} \; .
\end{equation}
Note that $\Delta_{CP}$ or $\Delta_T$ depends linearly on
the oscillating term $\sin F_{12}$,
therefore the length of the baseline suitable for
measuring $CP$ and $T$ asymmetries should satisfy the
condition $|\Delta m^2_{12}| \sim E/L$.
This requirement singles out
the large-angle MSW solution, which has $\Delta m^2_{\rm sun} \sim
10^{-5}$
to $10^{-4} ~ {\rm eV}^2$ and $\sin^2 2\theta_{\rm sun} \sim 0.65$
to $1$\cite{Valle}, among
three possible solutions to the solar neutrino problem.
The small-angle MSW solution is not favored; it does not give rise to
a relatively large magnitude of $\cal J$, which determines
the significance of practical $CP$- or $T$-violating signals.
The long wave-length vacuum oscillation requires $\Delta m^2_{\rm sun}
\sim 10^{-10}$ eV$^2$, too small to meet the realistic long-baseline
prerequisite.

In this talk I extend our previous hypothesis of lepton flavor
mixing\cite{FX96}, which arises naturally from the breaking of flavor
democracy for charged leptons and that of mass degeneracy for
neutrinos, to include $CP$ violation\cite{FX99L}. It is found that the
rephasing-invariant strength of $CP$ or $T$ violation can be as large
as one percent. The flavor mixing pattern remains
nearly bi-maximal, thus both atmospheric and solar neutrino
oscillations can well be interpreted. The consequences of
the model on $CP$ violation in the future long-baseline neutrino
experiments will also be discussed by taking the matter effects
into account. 

The phenomenological constraints obtained from various neutrino
oscillation experiments indicate that the mass differences in the
neutrino sector are tiny compared to those in the charged
lepton sector. One possible interpretation is that all three
neutrinos are nearly degenerate in mass. In this case one
might expect that the flavor mixing pattern of leptons differs
qualitatively from that of quarks, where both up and down
sectors exhibit a strong hierarchical structure in their mass
spectra and the observed mixing angles are rather small.
A number of authors have argued that the hierarchy of quark masses
and the smallness of mixing angles are related to each other,
by considering specific symmetry limits\cite{FX99}. One particular
way to proceed is to consider the limit of subnuclear democracy,
in which the mass matrices of both the up- and down-type quarks
are of rank one and have the structure
\begin{equation}
M_{\rm q} \; =\; \frac{c_{\rm q}}{3} \left ( \matrix{
1       & 1     & 1 \cr
1       & 1     & 1 \cr
1       & 1     & 1 \cr} \right )
\end{equation}
with q = u (up) or d (down) as well as $c_{\rm u} = m_t$
and $c_{\rm d} = m_b$.
Small departures from the democratic
limit lead to the flavor mixing and at the same time introduce
the masses of the second and first families. Specific symmetry
breaking schemes have been proposed in some literature
in order to calculate the flavor mixing angles in terms of
the quark mass eigenvalues\cite{FXReview}.

Since the charged leptons exhibit a similar hierarchical
mass spectrum as the quarks, it would be natural to consider
the limit of subnuclear democracy for the
$(e, \mu, \tau)$ system, i.e., the mass matrix takes the
form as Eq. (6). In the same
limit three neutrinos are degenerate in mass. Then we have\cite{FX96}
\begin{eqnarray}
M^{(0)}_l & = & \frac{c^{~}_l}{3} \left (\matrix{
1       & 1     & 1 \cr
1       & 1     & 1 \cr
1       & 1     & 1 \cr} \right ) \; ,
\nonumber \\
M^{(0)}_\nu & = & c_\nu \left (\matrix{
1       & 0     & 0 \cr
0       & 1     & 0 \cr
0       & 0     & 1 \cr} \right ) \; ,
\end{eqnarray}
where $c^{~}_l =m_\tau$ and $c_\nu =m_0$ measure the
corresponding mass scales.
If the three neutrinos are of the Majorana type,
$M^{(0)}_\nu$ could take a more general form
$M^{(0)}_\nu P_\nu$ with $P_\nu = {\rm Diag} \{ 1,
e^{i\phi_1}, e^{i\phi_2} \}$. As the Majorana phase matrix
$P_\nu$ has no effect on the flavor mixing and
$CP$-violating observables
in neutrino oscillations, it will be neglected in the subsequent
discussions.
Clearly $M^{(0)}_\nu$ exhibits an
$S(3)$ symmetry, while $M^{(0)}_l$ an
$S(3)_{\rm L} \times S(3)_{\rm R}$ symmetry.

One can transform the charged lepton mass matrix from the
democratic basis $M^{(0)}_l$ into the hierarchical basis
\begin{equation}
M^{(\rm H)}_l \; =\; c^{~}_l \left ( \matrix{
0       & 0     & 0 \cr
0       & 0     & 0 \cr
0       & 0     & 1 \cr } \right )
\end{equation}
by means of an orthogonal transformation, i.e.,
$M^{(\rm H)}_l = U M^{(0)}_l U^{\rm T}$, where
\begin{equation}
U \; =\; \left ( \matrix{
\frac{1}{\sqrt{2}}      & ~~ \frac{-1}{\sqrt{2}} ~      & 0 \cr
\frac{1}{\sqrt{6}}      & ~~ \frac{1}{\sqrt{6}} ~       &
\frac{-2}{\sqrt{6}} \cr
\frac{1}{\sqrt{3}}      & ~~ \frac{1}{\sqrt{3}} ~       &
\frac{1}{\sqrt{3}} \cr}
\right ) \; .
\end{equation}
We see $m_e = m_\mu =0$ from $M^{(\rm H)}_l$ and
$m_1 = m_2 =m_3 =m_0$ from $M^{(0)}_\nu$. Of course
there is no flavor mixing in this symmetry limit.

A simple real diagonal breaking of the flavor democracy
for $M^{(0)}_l$ and the mass degeneracy for $M^{(0)}_\nu$
may lead to instructive results for flavor mixing
in neutrino oscillations\cite{FX96,Tanimoto}.
To accommodate $CP$ violation, however, complex perturbative
terms are required.
Let us proceed with two different symmetry-breaking steps
in close analogy to the symmetry breaking discussed
previously for the quark mass matrices\cite{FP90,FH94}.

First, small real perturbations to the (3,3) elements of $M^{(0)}_l$
and $M^{(0)}_\nu$ are respectively introduced:
\begin{eqnarray}
\Delta M^{(1)}_l & = & \frac{c^{~}_l}{3} \left ( \matrix{
0       & 0     & 0 \cr
0       & 0     & 0 \cr
0       & 0     & \varepsilon^{~}_l \cr } \right ) \; ,
\nonumber \\
\Delta M^{(1)}_\nu & = & c_\nu \left ( \matrix{
0       & 0     & 0 \cr
0       & 0     & 0 \cr
0       & 0     & \varepsilon_\nu \cr } \right ) \; .
\end{eqnarray}
In this case the charged lepton mass matrix $M^{(1)}_l =
M^{(0)}_l + \Delta M^{(1)}_l$ remains symmetric under an
$S(2)_{\rm L}\times S(2)_{\rm R}$ transformation,
and the neutrino mass matrix
$M^{(1)}_\nu = M^{(0)}_\nu + \Delta M^{(0)}_\nu$ has
an $S(2)$ symmetry.
The muon becomes massive (i.e., $m_\mu \approx 2|\varepsilon^{~}_l|
m_\tau /9$), and the mass eigenvalue $m_3$ is no more degenerate
with $m_1$ and $m_2$ (i.e., $|m_3 - m_0| = m_0 |\varepsilon_\nu|$).
After the diagonalization of
$M^{(1)}_l$ and $M^{(1)}_\nu$, one finds that the 2nd and 3rd
lepton families have a definite flavor mixing angle
$\theta$. We obtain $\tan\theta \approx -\sqrt{2} ~$ if the
small correction of ${\cal O}(m_\mu/m_\tau)$ is neglected.
Neutrino oscillations at the atmospheric scale may arise
in $\nu_\mu \rightarrow \nu_\tau$ transitions with
$\Delta m^2_{32} = \Delta m^2_{31}
\approx 2m_0 |\varepsilon_\nu|$. The corresponding
mixing factor $\sin^2 2\theta \approx 8/9$ is in good agreement
with current data.

The symmetry breaking given in Eq. (10) for the charged lepton
mass matrix serves as a good illustrative example. One could
consider a more general case, analogous to the one for
quarks\cite{FP90}, to break the $S(3)_{\rm L}\times S(3)_{\rm R}$
symmetry of $M^{(0)}_l$ to an $S(2)_{\rm L} \times S(2)_{\rm R}$
symmetry. This would imply that $\Delta M^{(1)}_l$ takes the
form
\begin{equation}
\Delta M^{(1)}_l \; =\; \frac{c^{~}_l}{3}
\left ( \matrix{
0       & 0     & \varepsilon'_l \cr
0       & 0     & \varepsilon'_l \cr
\varepsilon'_l  & \varepsilon'_l        & \varepsilon^{~}_l \cr}
\right ) \; ,
\end{equation}
where $|\varepsilon^{~}_l| \ll 1$ and $|\varepsilon'_l| \ll 1$.
In this case the leading-order results obtained above, i.e.,
$\tan\theta \approx -\sqrt{2}$
and $\sin^2 2\theta \approx 8/9$, remain unchanged.

At the next step we introduce a complex symmetry breaking
perturbation, analogous to that for quark mass matrices\cite{Lehmann}, 
to the charged lepton mass
matrix $M^{(1)}_l$:
\begin{equation}
\Delta M^{(2)}_l \; = \; \frac{c^{~}_l}{3} \left ( \matrix{
0       & ~ -i\delta_l ~        & i\delta \cr
i\delta & ~ 0 ~         & -i\delta_l \cr
-i\delta_l      & ~ i\delta_l ~ & 0 \cr } \right ) \; .
\end{equation}
Transforming $M^{(2)}_l = M^{(1)}_l + \Delta M^{(2)}_l$ into
the hierarchical basis, we obtain
\begin{equation}
M^{\rm H}_l \; = \; c^{~}_l \left ( \matrix{
0       & -i\frac{1}{\sqrt{3}}\delta_l  & 0 \cr
i\frac{1}{\sqrt{3}}\delta_l     & \frac{2}{9}\varepsilon^{~}_l
& -\frac{\sqrt{2}}{9}\varepsilon^{~}_l \cr
0       & -\frac{\sqrt{2}}{9}\varepsilon^{~}_l
& 1 + \frac{1}{9}\varepsilon^{~}_l \cr}
\right ) \; .
\end{equation}
Note that $M^{\rm H}_l$, just like a variety of realistic
quark mass matrices\cite{FX99}, has texture zeros in the
(1,1), (1,3) and (3,1) positions.
The phases of its (1,2) and (2,1) elements are $\mp 90^{\circ}$,
which could lead to maximal $CP$ violation if the neutrino mass
matrix is essentially real.
For the latter we consider a simple perturbation 
to break the remaining mass degeneracy of
$M^{(1)}_\nu$:
\begin{equation}
\Delta M^{(2)}_\nu \; = \; c_\nu \left ( \matrix{
-\delta_\nu     & ~ 0   &  0 \cr
0       & ~ \delta_\nu  &  0 \cr
0       &  ~ 0  &  0 \cr } \right ) \; .
\end{equation}
We obtain $m_e \approx |\delta_l|^2 m^2_\tau /(27 m_\mu)$
and $m_{1,2} = m_0 (1 \mp \delta_\nu)$, respectively, from
$\Delta M^{(2)}_l$ and $\Delta M^{(2)}_\nu$.
The simultaneous diagonalization of
$M^{(2)}_l = M^{(1)}_l + \Delta M^{(2)}_l$ and
$M^{(2)}_\nu = M^{(1)}_\nu + \Delta M^{(2)}_\nu$
leads to a full $3\times 3$
flavor mixing matrix, which links neutrino mass eigenstates
$(\nu_1, \nu_2, \nu_3)$ to neutrino flavor eigenstates
$(\nu_e, \nu_\mu, \nu_\tau)$ in the following manner\cite{FX99L}:
\begin{equation}
V \; =\; U \; + \; i ~ \xi^{~}_V \sqrt{\frac{m_e}{m_\mu}}
\; + \; \zeta^{~}_V \frac{m_\mu}{m_\tau} \; ,
\end{equation}
where $U$ has been given in Eq. (9), and
\begin{eqnarray}
\xi^{~}_V & = &  \left ( \matrix{
\frac{1}{\sqrt{6}}      & ~ \frac{1}{\sqrt{6}} ~        &
\frac{-2}{\sqrt{6}} \cr
\frac{1}{\sqrt{2}}      & ~ \frac{-1}{\sqrt{2}} ~       & 0 \cr
0       & ~ 0 ~ & 0 \cr} \right ) \; ,
\nonumber \\
\zeta^{~}_V & = & \left ( \matrix{
0       & 0     & 0 \cr
\frac{1}{\sqrt{6}}      & \frac{1}{\sqrt{6}}    & \frac{1}{\sqrt{6}} \cr
\frac{-1}{\sqrt{12}}    & \frac{-1}{\sqrt{12}}  & \frac{1}{\sqrt{3}}
\cr} \right ) \; .
\end{eqnarray}
In comparison with the previous result\cite{FX96},
the new feature of this lepton mixing scenario is that
the term multiplying $\xi^{~}_V$ becomes imaginary. Therefore $CP$ or
$T$ violation has been incorporated.

The complex symmetry breaking perturbation given in Eq. (12) is
certainly not the only one which can be considered for $M^{(1)}_l$.
A number of other interesting patterns of $\Delta M^{(2)}_l$,
which result in the same flavor mixing matrix as that given in Eq. (15),
have been discussed by us in a recent paper\cite{FX99L}.
Here we remark that Hermitian perturbations of the same
forms as given in Eqs. (11) and (12) have been used to break the
flavor democracy of quark mass matrices and to generate $CP$
violation\cite{FP90,Lehmann}. The key point of this similarity
between the charged lepton and quark mass matrices is that both of
them have the strong mass hierarchy and might have the same
dynamical origin or a symmetry relationship.

The flavor mixing matrix $V$ can in general be parametrized
in terms of three Euler angles and one $CP$-violating phase
\footnote{For
neutrinos of the Majorana type, two additional
$CP$-violating phases may enter. But they are irrelevant
to neutrino oscillations and can be neglected for our present
purpose.}.
A suitable parametrization reads as follows\cite{FX97}:
\begin{eqnarray}
V & = & \left ( \matrix{
c^{~}_l & s^{~}_l       & 0 \cr
-s^{~}_l        & c^{~}_l       & 0 \cr
0       & 0     & 1 \cr } \right )  \left ( \matrix{
e^{-i\phi}      & 0     & 0 \cr
0       & c     & s \cr
0       & -s    & c \cr } \right )  \left ( \matrix{
c_{\nu} & -s_{\nu}      & 0 \cr
s_{\nu} & c_{\nu}       & 0 \cr
0       & 0     & 1 \cr } \right )  \nonumber \\ \nonumber \\
& = & \left ( \matrix{
s^{~}_l s_{\nu} c + c^{~}_l c_{\nu} e^{-i\phi} & ~~~~
s^{~}_l c_{\nu} c - c^{~}_l s_{\nu} e^{-i\phi} ~~~~ &
s^{~}_l s \cr
c^{~}_l s_{\nu} c - s^{~}_l c_{\nu} e^{-i\phi} &
c^{~}_l c_{\nu} c + s^{~}_l s_{\nu} e^{-i\phi}   &
c^{~}_l s \cr - s_{\nu} s       & - c^{~}_\nu s & c \cr } \right ) \; ,
\end{eqnarray}
in which $s^{~}_l \equiv \sin\theta_l$, $s_{\nu} \equiv
\sin\theta_{\nu}$,
$c \equiv \cos\theta$, etc. The three mixing angles can all be
arranged to lie in the first quadrant, while the $CP$-violating
phase may take values between $0$ and $2\pi$.
It is straightforward to obtain
${\cal J} = s^{~}_l c^{~}_l s_\nu c_\nu s^2 c \sin\phi$.
Numerically we find
\begin{equation}
\theta_l \approx 4^{\circ} \; , ~~~~~
\theta_\nu \approx 45^{\circ} \; , ~~~~~
\theta \approx 52^{\circ} \; , ~~~~~
\phi \approx 90^{\circ} \;
\end{equation}
from Eq. (15).
The smallness of $\theta_l$ is a natural consequence of the
mass hierarchy in the charged lepton sector, just as the
smallness of $\theta_{\rm u}$ in quark mixing\cite{FX99}.
On the other hand, both $\theta_\nu$ and $\theta$ are too large
to be comparable with the corresponding quark mixing angles
$\theta_{\rm d}$ and $\theta$\cite{FX99},
reflecting the qualitative difference between quark and lepton
flavor mixing phenomena. It is worth emphasizing that the
leptonic $CP$-violating phase $\phi$ takes a special value
($\approx 90^{\circ}$) in our model. The same possibility
is also favored for the quark mixing phenomenon in a variety of
realistic mass matrices\cite{FX95}.
Therefore maximal leptonic $CP$ violation, in the sense that
the magnitude of ${\cal J}$ is maximal for the fixed values
of three flavor mixing angles, appears naturally as in the
quark sector.

Some consequences of this lepton mixing scenario
can be drawn as follows:

(a) The mixing pattern in Eq. (15), after neglecting small
corrections from the charged lepton masses, is quite similar to that
of the pseudoscalar mesons $\pi^0$, $\eta$ and $\eta'$ in QCD in
the limit of the chiral $SU(3)_{\rm L} \times SU(3)_{\rm R}$
symmetry\cite{FH94,Ringberg}:
\begin{eqnarray}
\pi^0 & = & \frac{1}{\sqrt{2}} \left (
|\bar{u}u\rangle - |\bar{d}d\rangle \right ) \; ,
\nonumber \\
\eta & = & \frac{1}{\sqrt{6}} \left (
|\bar{u}u\rangle + |\bar{d}d\rangle - 2 |\bar{s}s\rangle
\right ) \; ,
\nonumber \\
\eta' & = & \frac{1}{\sqrt{3}} \left (
|\bar{u}u\rangle + |\bar{d}d\rangle + |\bar{s}s\rangle
\right ) \; .
\end{eqnarray}
Some preliminary theoretical attempts towards deriving the flavor
mixing matrix $V\approx U$ have been reviewed elsewhere\cite{FXReview}.

(b) The $V_{e3}$ element, of magnitude
\begin{equation}
|V_{e3}| \; =\; \frac{2}{\sqrt{6}} \sqrt{\frac{m_e}{m_\mu}} \;\; ,
\end{equation}
is naturally suppressed in
the symmetry breaking scheme outlined above.
A similar feature appears in the $3\times 3$ quark flavor mixing
matrix, i.e., $|V_{ub}|$ is the smallest among the
nine quark mixing elements. Indeed the smallness of $V_{e3}$
provides a necessary condition for the decoupling of
solar and atmospheric neutrino oscillations, even though neutrino
masses are nearly degenerate. The effect of small but nonvanishing
$V_{e3}$ will manifest itself in long-baseline $\nu_\mu
\rightarrow \nu_e$ and $\nu_e \rightarrow \nu_\tau$ 
transitions\cite{FX96}.

(c) The flavor mixing between the 1st and 2nd lepton families
and that between the 2nd and 3rd lepton families are nearly
maximal. This property, together with the natural smallness
of $|V_{e3}|$, allows a satisfactory interpretation of the
observed large mixing
in atmospheric and solar neutrino oscillations. We obtain
\footnote{In calculating $\sin^2 2\theta_{\rm sun}$ we have taken
the ${\cal O}(m_e/m_\mu)$ correction to $V$ into account.}
\begin{eqnarray}
\sin^2 2\theta_{\rm sun} & = & 1 - \frac{4}{3} \frac{m_e}{m_\mu}
\; , \nonumber \\
\sin^2 2\theta_{\rm atm} & = & \frac{8}{9} +
\frac{8}{9} \frac{m_\mu}{m_\tau} \;
\end{eqnarray}
to a high degree of accuracy. Explicitly
$\sin^2 2\theta_{\rm sun} \approx 0.99$ and $\sin^2 2\theta_{\rm atm}
\approx 0.94$, favored by current data\cite{SK}. It is obvious that the
model is fully consistent with the vacuum oscillation solution to
the solar neutrino problem and might also be
able to incorporate the large-angle MSW solution
\footnote{A slightly
different symmetry-breaking pattern of the neutrino mass
matrix\cite{Tanimoto00}, which involves four free parameters,
allows the magnitude of $\sin^2 2\theta_{\rm sun}$ to be smaller
and also consistent with the large-angle MSW solution.}.

(d) It is worth remarking that our lepton mixing pattern has no conflict
with current constraints on the neutrinoless double beta
decay\cite{Beta}, if neutrinos are of the Majorana type.
In the presence of $CP$ violation, the effective
mass term of the $(\beta\beta)_{0\nu}$ decay can be written as
\begin{equation}
\langle m\rangle_{(\beta\beta)_{0\nu}} \; = \;
\sum^3_{i=1} \left (m_i ~ \tilde{V}^2_{ei} \right ) \; ,
\end{equation}
where $\tilde{V} = VP_\nu$
and $P_\nu = {\rm Diag}\{1, e^{i\phi_1}, e^{i\phi_2} \}$ is the
Majorana phase matrix. If the unknown phases are taken to be
$\phi_1 =\phi_2 =90^{\circ}$ for example, then one arrives at
\begin{equation}
\left | \langle m \rangle_{(\beta\beta)_{0\nu}} \right | \; = \;
\frac{2}{\sqrt{3}} \sqrt{\frac{m_e}{m_\mu}}
~ m_i \; ,
\end{equation}
in which $m_i \sim 1 - 2$ eV (for $i=1,2,3$) as required by the
near degeneracy of three neutrinos in our model
to accommodate the hot dark matter of the universe.
Obviously $|\langle m\rangle_{(\beta\beta)_{0\nu}}|
\approx 0.08 m_i \leq 0.2$ eV, the
latest bound of the $(\beta\beta)_{0\nu}$ decay\cite{Beta}.

(e) The rephasing-invariant strength of $CP$ violation in
this scheme is given as\cite{FX99L}
\begin{equation}
{\cal J} \; = \; \frac{1}{3\sqrt{3}} \sqrt{\frac{m_e}{m_\mu}}
\left ( 1 + \frac{1}{2} \frac{m_\mu}{m_\tau} \right ) \; .
\end{equation}
Explicitly we have ${\cal J}
\approx 1.4\%$. The large magnitude of $\cal J$ for lepton mixing
turns out to be very non-trivial, as the same quantity for quark mixing
is only of the ${\cal O}(10^{-5})$ level\cite{FX99,FX95}.
If the mixing pattern under discussion
were in no conflict with the large-angle MSW solution to the
solar neutrino problem, then
the $CP$- and $T$-violating signals
$\Delta_{CP} = \Delta_T \propto -16 {\cal J} \approx -0.2$
could be significant enough to be measured from the asymmetry
between $P(\nu_\mu \rightarrow \nu_e)$ and
$P(\bar{\nu}_\mu \rightarrow \bar{\nu}_e)$ or that
between $P(\nu_\mu \rightarrow \nu_e)$ and $P(\nu_e \rightarrow
\nu_\mu)$ in the long-baseline
neutrino experiments. In the leading-order approximation
we arrive at
\begin{eqnarray}
{\cal A} & = & \frac{P(\nu_\mu \rightarrow \nu_e) ~ - ~ P(\bar{\nu}_\mu
\rightarrow \bar{\nu}_e)}{P(\nu_\mu \rightarrow \nu_e) ~ + ~
P(\bar{\nu}_\mu \rightarrow \bar{\nu}_e)}
\nonumber \\ \nonumber \\
& = & \frac{P(\nu_\mu \rightarrow \nu_e) ~ - ~ P(\nu_e \rightarrow
\nu_\mu)}
{P(\nu_\mu \rightarrow \nu_e) ~ + ~ P(\nu_e \rightarrow \nu_\mu)}
\nonumber \\ \nonumber \\
& = & \frac{\displaystyle -\frac{8}{\sqrt{3}} \sqrt{\frac{m_e}{m_\mu}}}
{\displaystyle \frac{16}{3} \frac{m_e}{m_\mu}
+ \left (\frac{\sin F_{12}}{\sin F_{23}} \right )^2} ~ \sin F_{12} \; .
\end{eqnarray}
The asymmetry ${\cal A}$ depends linearly on
the oscillating term $\sin F_{12}$, which is associated essentially with
the solar neutrino anomaly.

Note that ${\cal A}$ signals $CP$ or $T$ violation solely in vacuum. For
most of the proposed long-baseline neutrino experiments the
earth-induced
matter effects on neutrino oscillations are non-negligible and should
be carefully handled. In matter the effective Hamiltonian of
neutrinos can be written as\cite{Wolfenstein}
\begin{equation}
{\cal H}_\nu \; = \; \frac{1}{2E} \left [ V \left (\matrix{
m^2_1     & 0      & 0 \cr
0         & m^2_2  & 0 \cr
0         & 0      & m^2_3 \cr} \right ) V^{\dagger}
~ + ~ A \left ( \matrix{
1     & 0     & 0 \cr
0     & 0     & 0 \cr
0     & 0     & 0 \cr} \right ) \right ] \; ,
\end{equation}
where $A= 2\sqrt{2} G_{\rm F} N_e E$ describes the charged-current 
contribution to the $\nu_e e^-$ forward scattering,
$N_e$ is the background density of electrons, and $E$ stands for
the neutrino beam energy. The neutral-current contributions are
universal for $\nu_e$, $\nu_\mu$ and $\nu_\tau$ neutrinos,
leading only to an overall unobservable phase and have been
neglected. The effective Hamiltonian responsible for antineutrinos
propagating in matter, defined as ${\cal H}_{\bar \nu}$, can 
be obtained from ${\cal H}_\nu$ with the replacements 
$V \rightarrow V^*$ and $A \rightarrow -A$.
Using ${\cal H}_\nu$ and ${\cal H}_{\bar \nu}$
one can derive the effective neutrino mass eigenvalues and
the effective flavor mixing matrices in matter\cite{Xing00}.

For simplicity we only present the numerical results of the
matter-corrected $CP$- and $T$-violating asymmetries in the
assumption of the baseline length $L = 732$ km, i.e., a
neutrino source at Fermilab pointing toward the Soudan mine
in Minnesota or that at CERN toward the Gran Sasso underground
laboratory in Italy. The inputs include the flavor mixing and
$CP$-violating parameters obtained in Eq. (15) as well as
the typical neutrino mass-squared differences
$\Delta m^2_{21} = 5 \cdot 10^{-5} ~ {\rm eV}^2$ and
$\Delta m^2_{32} = 3 \cdot 10^{-3} ~ {\rm eV}^2$.
Assuming a constant earth density profile, one has
$A \approx 2.28 \cdot 10^{-4} ~ {\rm eV}^2 E/[{\rm GeV}]$\cite{Barger99}. 
The behaviors of
the $CP$ and $T$ asymmetries changing with the beam energy $E$
in the range $3 ~{\rm GeV} \leq E \leq 20 ~{\rm GeV}$
are shown in Fig. 1. 
Clearly the vacuum asymmetry ${\cal A}$ can be of ${\cal O}(0.1)$.
The matter-induced correction to the $T$-violating asymmetry
is negligibly small for the experimental conditions under consideration.
In contrast, the matter effect on the $CP$-violating asymmetry
cannot be neglected, although it is unable to
fake the genuine $CP$-violating signal.
\begin{figure}[t]
\epsfig{file=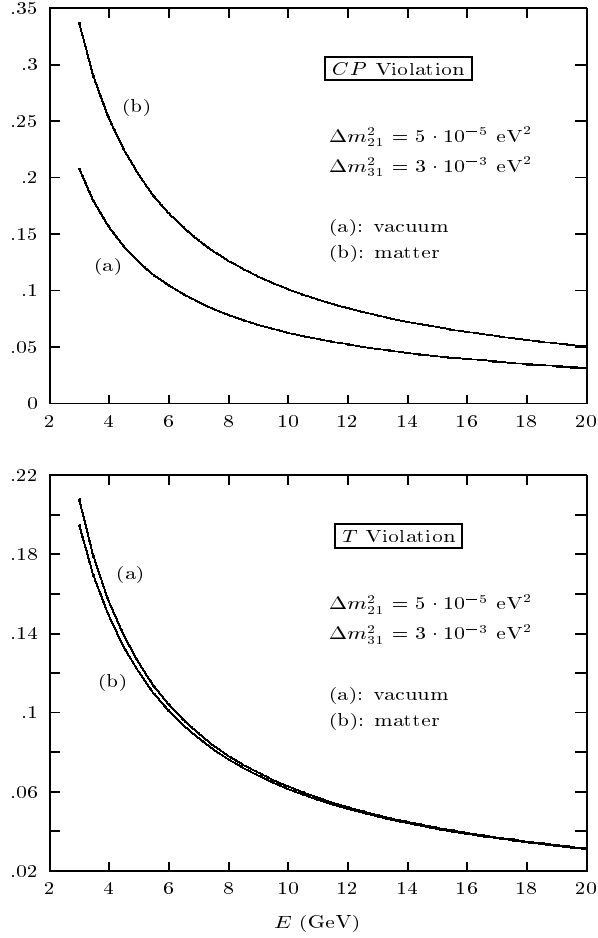,bbllx=4cm,bblly=-4cm,bburx=20cm,bbury=27cm,%
width=13.5cm,height=21cm,angle=0,clip=}
\vspace{-8.1cm}
\caption{Illustrative plots for the $CP$-violating asymmetries
between $\nu_\mu \rightarrow \nu_e$ and
$\bar{\nu}_\mu \rightarrow \bar{\nu}_e$ transitions as well as
the $T$-violating
asymmetries between $\nu_\mu \rightarrow \nu_e$ and
$\nu_e \rightarrow \nu_\mu$ transitions, in vacuum
and in matter, changing with the neutrino beam energy $E$.}
\end{figure}

If the upcoming data appeared to rule out the consistency between our
model and the large-angle MSW solution to the solar neutrino problem,
then it would be quite difficult
to test the model itself from its prediction for
large $CP$ and $T$ asymmetries in any realistic long-baseline
experiment.

In summary, I have extended our previous model of the
nearly bi-maximal lepton flavor mixing to incorporate large
$CP$ violation. The new model remains favored by current
data on atmospheric and solar neutrino oscillations, and
it predicts significant $CP$- and $T$-violating effects
in the long-baseline neutrino experiments. I expect that
more data from the Super-Kamiokande and other neutrino
experiments could soon provide stringent tests of the
existing lepton mixing models and give useful hints
towards the symmetry or dynamics of lepton mass generation.

\end{document}